\documentstyle[12pt,a4,graphicx]{article}

\newcommand{\be}{\begin{equation}}
\newcommand{\ee}{\end{equation}}
\newcommand{\ba}{\begin{eqnarray}}
\newcommand{\ea}{\end{eqnarray}}

\newcommand{\re}{\mbox{Re}\,}
\newcommand{\im}{\mbox{Im}\,}
\newcommand{\tr}{\mathrm{tr}}

\begin{document}
\begin{titlepage}
\begin{flushright}
CAFPE-23/03\\
LU TP 03-43\\
UG-FT-153/03\\
hep-ph/0309216\\
\end{flushright}
\vspace{2cm}
\begin{center}

{\large\bf Hadronic Matrix Elements for Kaons
\footnote{Work supported in part 
by NFR, Sweden, 
by MCYT, Spain (Grants No. FPA2000-1558 and HF2001-0116), 
by Junta de Andaluc\'{\i}a, (Grant No. FQM-101),
and by the European Union RTN Contract No. HPRN-CT-2002-00311 (EURIDICE).
Invited talk given by J.P.
at ``X International Conference on Quantum Chromodynamics
 (QCD '03)'', 2-9 July 2003, Montpellier, France.}}\\
\vfill
{\bf Johan Bijnens$^{a)}$, Elvira G\'amiz$^{b)}$
 and Joaquim Prades$^{b)}$}\\[0.5cm]
$^{a)}$ Department of Theoretical Physics 2, Lund University \\
S\"olvegatan 14A, S-22362 Lund, Sweden.\\[0.5cm]

$^{b)}$ Centro Andaluz de F\'{\i}sica de las Part\'{\i}culas
Elementales (CAFPE) and Departamento de
 F\'{\i}sica Te\'orica y del Cosmos, Universidad de Granada \\
Campus de Fuente Nueva, E-18002 Granada, Spain.\\[0.5cm]
\end{center}
\vfill
\begin{abstract}
\noindent
We review some work done by us calculating
matrix elements for Kaons. Emphasis is put on the
matrix elements which are 
relevant to predict non-leptonic Kaon CP violating
observables.
In particular, we recall our results for the 
$\hat B_K$ parameter which governs the $\overline{K^0}-K^0$
mixing and update
our results for $\varepsilon'_K$ including estimated all-higher-order
CHPT corrections and the new results from
recent analytical calculations of the $\Delta I=3/2$ component.
 Some comments on future prospects on
calculating matrix elements for Kaons are also added.
\end{abstract}
\vfill
September 2003
\end{titlepage}

\section{Outline}
Here we review some work done by us in 
\cite{BK,DeltaI=1/2,epsprime,scheme,Q7Q8}
devoted to the calculation of matrix elements of
Kaons.  The main motivation for these calculations
is the reduction of the hadronic uncertainty
in those matrix elements 
which  are necessary to test the Standard Model
and unveil beyond the Standard Model physics.

We will limit our discussion to matrix elements
related to CP-violating effects involving Kaons:
namely, the $B_K$-parameter governing
the {\em indirect} CP-violation in the
 $\overline{K^0}-K^0$ mixing \cite{BK,scheme} and
the {\em direct} CP-violating parameter $\varepsilon_K'$
\cite{epsprime,Q7Q8}.
We update $\varepsilon_K'$ by including
the non-Final State Interaction (non-FSI) contributions,
the new full isospin breaking corrections \cite{CPEN03}
as well as new results for some of the inputs used.
We also use the results from the analytical results
for the $\Delta I=3/2$ component \cite{Q7Q8} to substitute the ones
used in \cite{epsprime}, these two are, incidentally, 
in numerical agreement.

We don't discuss
the light-by-light hadronic contributions to the muon $g-2$
\cite{g-2} nor the
matrix elements of Kaons needed to quantify the
electromagnetic mass difference \cite{EMmass}. 
Both calculations  were done using techniques similar to 
those discussed here and can be improved along the
lines discussed in the last section.

\section{Motivation and Notation}
\label{motiv}

We study here matrix elements related to 
two aspects of  CP violation in Kaon physics; namely,

a) The $\hat B_K$ parameter, defined as 
[$C(\nu)$  is a Wilson coefficient]
\ba
&& \hskip-0.7cm C(\nu) \, \langle {\overline K^0} |
\int {\rm d}^4 x \, {\cal Q}_{\Delta S=2}(x)
|{K^0} \rangle \equiv \frac{16}{3} \, \hat B_K
\, f_K^2 m_K^2 \, ; \nonumber \\
&&\hskip-0.5cm {\cal Q}_{\Delta S=2}(x) = 4 L^\mu(x) L_\mu(x) 
\hskip0.5cm {\rm and} \\
&&\hskip-0.5cm 
2 L^\mu(x) = [\overline s \gamma^\mu (1-\gamma_5) d ](x) \, .
\nonumber
\ea
which enters in the {\em indirect} CP violating
parameter $\varepsilon_K$. The $\hat B_K$ parameter is an important
input for the analysis of one of the Cabibbo-Kobayashi-Maskawa 
(CKM) unitarity triangles --see \cite{UT} for more information--

and b) The {\em direct} CP violation in $K\to \pi\pi$
decays parameterized through $\varepsilon'_K$, defined as
\ba
&&\hskip-0.7cm
\frac{\sqrt{2}\varepsilon_K'}{\varepsilon_K}=
\frac{A\left[K_L\!\to\!(\pi\pi)_{I=2}\right]}
{A\left[K_L\!\to\!(\pi\pi)_{I=0}\right]}-
\frac{A\left[K_S\!\to\!(\pi\pi)_{I=2}\right]}
{A\left[K_S\!\to\!(\pi\pi)_{I=0}\right]}
 .
\nonumber
\ea
In the isospin symmetry limit, $K\to\pi\pi$ invariant amplitudes
can be decomposed into definite isospin amplitudes as
$[A\equiv-i T]$
\ba
i \, A[K^0\to \pi^0\pi^0] &\equiv& {\frac{a_0}{\sqrt 3}} \, 
 e^{i\delta_0}
-\frac{ 2 \, a_2}{\sqrt 6} \, e^{i\delta_2} \, , \nonumber \\
i \, A[K^0\to \pi^+\pi^-] &\equiv& {\frac{a_0}{\sqrt 3}} \, 
 e^{i\delta_0}
+\frac{a_2}{\sqrt 6} \, e^{i\delta_2}\,  
\ea
with $\delta_0$ and $\delta_2$ the
FSI phases and
under very reasonable approximations, one can get
\be
\varepsilon'_K
\simeq \frac{i e^{i(\delta_2-\delta_0)}}{\sqrt 2} \frac{\re a_2}{\re a_0}
\left[-\frac{\im a_0}{\re a_0}+\frac{\im a_2}{\re a_2} \right] \, .
\ee

To lowest order (LO) in Chiral Perturbation Theory (CHPT), i.e. 
order $e^0 p^2$ and $e^2 p^0$, strong and electromagnetic
interactions between  $\pi$, $K$, $\eta$ and external sources are 
described by
\be
\label{lagstrong}
{\cal L}^{(2)} =
\frac{F_0^2}{4}\tr\left(u_\mu u^\mu+\chi_+\right)
+ e^2 \widetilde C_2 \tr\left(Q U Q U^\dagger\right)
\ee
with $U = u^2 = \exp(\lambda^a\pi^a/F_0)$ and 
$u_\mu = i u^\dagger  (D_\mu U) u^\dagger$.  
$\lambda^a$ are the Gell-Mann matrices and the $\pi^a$
are the pseudo-scalar-mesons $\pi$, $K$, and $\eta$.
$Q=\mbox{diag}(2/3,-1/3,-1/3)$ is the light-quark-charge matrix and
$\chi_+ = u^\dagger\chi u^\dagger + u \chi^\dagger u$ and
$\chi = 2 B_0 \, \mbox{diag}(m_u,m_d,m_s)$ collects the light-quark 
masses. To this order, $f_\pi=F_0=$ 87 MeV is the pion decay coupling 
constant.  Introductions to CHPT can be found in \cite{CHPTlectures}.

To the same order in CHPT, the chiral Lagrangians
describing $|\Delta S|=2$  and $|\Delta S|=1$ transitions 
are respectively
\ba
\label{lagdS2}
&&\hskip-1cm{\cal L}^{(2)}_{|\Delta S|=2}=
\widehat C F_0^4 G_{\Delta S=2}
 \tr\left (\Delta_{32} u^\mu \right)
\tr \left( \Delta_{32} u_\mu\right) +\mbox{h.c.}
\nonumber \\
\ea
and 
\ba
\label{lagdS1} 
{\cal L}^{(2)}_{|\Delta S|=1}&=&
C F_0^6 e^2 G_E \,  \tr\left(\Delta_{32}u^\dagger Q u  \right)
\nonumber \\&&\hskip-2.0cm+
 C F_0^4 \Bigg[G_8\tr\left(\Delta_{32}u_\mu u^\mu\right)
+G_8^\prime\tr\left(\Delta_{32}\chi_+\right) 
\nonumber \\&&\hskip-2.0cm +
G_{27}t^{ij,kl}\tr\left(\Delta_{ij}U_\mu\right)
  \tr\left(\Delta_{kl}u^\mu\right)\Bigg] +\mbox{h.c.} \ .
\ea
Here $\Delta_{ij} = u\lambda_{ij}u^\dagger$ and
$(\lambda_{ij})_{ab} = \delta_{ia}\delta_{jb}$.
\be
C= -\frac{3}{5} \frac{G_F}{\sqrt 2} \, V_{ud} V_{us}^* \approx
-1.07 \times 10^{-6} \, {\rm GeV}^{-2} \, . 
\ee
The normalization $\widehat C$ is a known function
of  the  $W$-boson, top and charm quark masses and
of the CKM matrix elements.  The SU(3) $\times$ SU(3) tensor
$t^{ij,kl}$ can be found in \cite{BPP98}.

In the Standard Model (SM), 
$\im  G_{27}$ vanishes and
$\im  G_8$ and $\im  G_E$ are proportional
to CP-violating phases --to $\im  \tau$ with
 $\tau \equiv -\lambda_t /\lambda_u$ and $\lambda_i\equiv
V_{id} V_{is}^*$ where $V_{ij}$ are CKM matrix elements. 

At this order one gets
\be
\hat B_K^{\chi} = \frac{3}{4} \, G_{\Delta S=2} \, ,
\ee
\be
\label{LOima0}
\left(\frac{\re a_0}{\re a_2}\right)^{(2)} \simeq {\sqrt 2}  \, 
\frac{ 9 \re G_8 + G_{27}}{10 G_{27}} \, ,
\ee
\be
\label{LOrea0}
\left(\frac{\im a_0}{\re a_0}\right)^{(2)} 
\simeq \frac{\im G_8}{\re G_8 + G_{27}/9} \, , 
\ee
and
\be 
\label{LOima2}
\left(\frac{\im a_2}{\re a_2}\right)^{(2)}
 \simeq -\frac{3}{5} \frac{F_0^2}{m_K^2-m_\pi^2}
\frac{\im (e^2 G_{E})}{G_{27}} \, .
\ee
where we have disregarded some tiny electroweak corrections
proportional to $\re(e^2 G_E)$.

 The large number of colors ($N_c$) predictions for the couplings
in (\ref{lagdS2})  and (\ref{lagdS1}) come from factorisable diagrams,
one gets
\ba
G_{\Delta S=2} = G_{8}= G_{27}= 1 \, ; \hskip0.5cm
{\rm and}  \hskip0.5cm G_E = 0 \, .
\ea
Our goal in \cite{BK,DeltaI=1/2,epsprime,scheme,Q7Q8} 
was to calculate the next-to-leading order (NLO)
in $1/N_c$ contributions to these couplings
as well as the chiral corrections to $\hat B_K$.
Due to the lack of space, we will not go into details
about the technique used to calculate hadronic matrix
elements in those references.
All the details on the X-boson method
and on the short-distance matching were given there.
Just to comment here that, in general, we compute a 
two-point function 
\ba
\label{twopoint}
\lefteqn{
{\bf \Pi}_{ij}(q^2) \equiv}&&\\
&& i \int {\rm d}^4 x \, e^{iq \cdot x}
\, \langle 0 | T \left( P_{i}^\dagger(0) P_j(x) e^{i {\bf \Gamma_{LD}}}
 \right)| 0 \rangle
\nonumber
\ea
in the presence of the long-distance effective
action of the Standard Model ${\bf \Gamma_{\rm LD}}$. 
The pseudo-scalar sources $P_i(x)$ have the appropriate quantum
numbers to describe $K\to\pi$ transitions. 
The effective action ${\bf \Gamma_{LD}}$
reproduces the physics of the SM at low energies by the exchange of 
colorless heavy  X-bosons.  To obtain it we make  a
short-distance matching analytically, which takes into account exactly the
short-distance scale and scheme dependence.
 We are left with the couplings 
of the  X-boson long-distance effective action
completely fixed in terms of the Standard Model ones.
 This action is regularized with a four-dimensional
 cut-off, $\mu_C$. Our X-boson
effective action has the technical advantage to separate the short-distance
of the two-quark currents or densities from the purely four-quark
short-distance which is always only logarithmically divergent and 
regularized by the X boson mass in our approach.
 The cut-off $\mu_C$ only appears in the short-distance of  the
two-quark currents or densities and can be thus taken into account 
exactly.

 Taylor expanding the two-point function  (\ref{twopoint})
in $q^2$ and quark masses one can extract  the CHPT couplings
$G_{\Delta S=2}$, $G_8$, $G_{27}$, $\cdots$ and make the
predictions of the physical quantities at lowest order. One can also
go further and extract the NLO CHPT weak counterterms needed
for instance in the isospin breaking corrections or in the
rest of NLO CHPT corrections.

After following the procedure sketched above we are able to write
$\hat B_K$, $\re G_8$, $\im G_8$, and $G_{27}$ 
as some known effective coupling  \cite{epsprime,scheme}
$|g_i|^2(M_X,\mu_C,\cdots)$ times
\be
\label{PPAB}
\int^\infty_0 {\rm d} Q^2 \,  \frac{Q^{2}}{Q^2+M_X^2} \, 
{\bf \Pi}_{PPAB}(Q^2,q^2)
\ee
where ${\bf \Pi}_{PPAB}(Q^2,q^2)$ is a four-point function
with $AB$  being either $L^\mu L_\mu$ or $L^\mu R_\mu$;
$L^\mu$ and $R^\mu$ are  left and right currents, respectively, 
and $Q=|p_X^E|$
is the X-boson momentum in Euclidean space. 
The case of $\im (e^2 G_E)$ can be written as~\cite{Q7Q8}
\be
\label{LR}
\int^\infty_0 {\rm d} Q^2 \, \frac{Q^{4}}{Q^2+M_X^2} \, 
{\bf \Pi}_{LR}(Q^2) \, .
\ee
Therefore in this case
one can use data as done in 
\cite{Q7Q8,CDGM03,NAR01}. In \cite{KPR01}, a Minimal
Hadronic Approximation to large $N_c$ was used to saturate
the relevant two-point function.

We split the 0 to $\infty$ integration  in (\ref{PPAB}) and 
(\ref{LR}) into  a long-distance (LD) piece (from 
0 to $\mu$) and a short-distance piece (SD) (from $\mu$ to $\infty$).
The SD piece we do using the Operator Product Expansion (OPE) in QCD
and the result is thus model independent.
For the long-distance piece we used  data for ${\bf \Pi}_{LR}$
and, as a first step,
the ENJL model \cite{ENJL} for the four-point Green's functions. 
 The good features and drawbacks
of the ENJL model we used have been raised several times 
in \cite{epsprime,g-2}. The most important drawback is that
it does not contain all QCD constraints \cite{PPR98}.
 There is work in progress to substitute it
 by a  ladder resummation inspired
hadronic model that in addition to the
good features of the ENJL model (contains
some short-distance QCD
constraints, CHPT up to order $p^4$, 
good phenomenology, $\cdots$) includes large $N_c$ QCD as
well as more short-distance QCD constraints, see last section.

\section{Results for $\hat B_K$}

Here  we only recall our results in \cite{BK,scheme}.
\begin{figure}[thb]
\begin{center}
\includegraphics[height=0.45\textwidth]{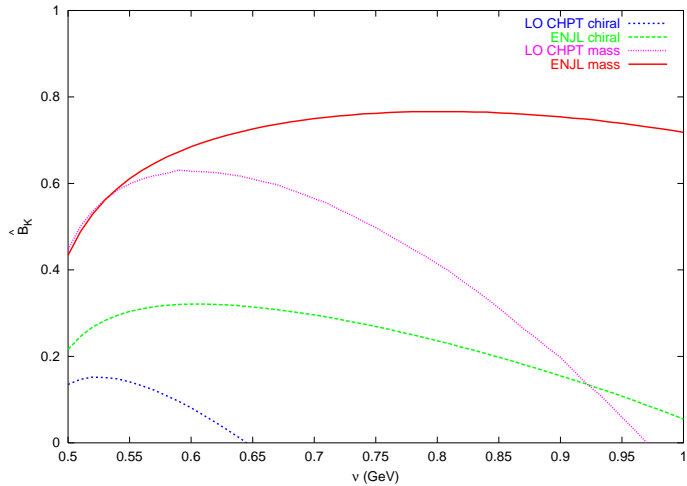}
\caption{\label{BK}  Matching of the short-distance scale
for  our $\hat B_K$ prediction.  From  down to up curves:
using CHPT at LO in the chiral limit,
 using ENJL in the chiral limit, using CHPT at LO with physical masses,
using ENJL with physical masses. Notice the quality of this last
curve.}
\end{center}
\end{figure}
We got good numerical scale matching for $\hat B_K$
for the 
real case and not so good for the $\chi$-limit $\hat B_K^\chi$,
 see Figure \ref{BK}.
Our results are also analytically scheme independent \cite{scheme}.
\be
\hat B_K^\chi = \frac{3}{4} (0.43 \pm 0.20) = 0.32 \pm 0.15
\ee
in the chiral limit and
\be
\hat B_K = 0.77 \pm 0.10 
\ee
for the case of real pion and Kaon masses.

\section{Results for $\varepsilon'_K$}

In \cite{epsprime} we made a prediction for
$\varepsilon'_K$ which we would like to update now.
The new things we would like to input are the non-FSI
corrections which after the work in \cite{BPP98,PPS01,CG02,BDP03}
 are known.
We also use the recent complete isospin breaking
result of \cite{CPEN03} and
the new analytical results for $\im (e^2G_E)$ \cite{Q7Q8,CDGM03,NAR01}.
Our result in \cite{epsprime} did contain the FSI corrections
but not the non-FSI which were unknown at that time.
 
In \cite{Q7Q8,CDGM03,NAR01} 
there are recent calculations
of $\im (e^2G_E)$ using dispersion relations.
 The results there are valid to all orders
in $1/N_c$ and NLO in $\alpha_S$. They are obtained using
the hadronic tau data collected by ALEPH \cite{ALEPH} and 
OPAL \cite{OPAL} at LEP.
The agreement between them is quite good and their
results can be summarized in
\be
\label{Q8}
\im (e^2 G_E) = - (4.0\pm0.9) \, 
\left( \frac{87 {\rm MeV}}{F_0}\right)^6\,
 \im \tau \, .
\ee
In the Standard Model
\be
\label{tau}
\im \tau = -(6.05 \pm 0.50) \times 10^{-4} \, .
\ee

In \cite{KPR01}, they used a Minimal Hadronic Approximation
(MHA)
to large $N_c$ QCD to calculate $\im(e^2 G_E)$ with the result
\be
\im (e^2 G_E) = -(6.7\pm2.0) \, 
\left( \frac{87 {\rm MeV}}{F_0}\right)^6\,
\im \tau \, ,
\ee
which is also in agreement with though somewhat larger
than (\ref{Q8}). There are also lattice results for
$\im(e^2 G_E)$ both using domain-wall fermions \cite{domain}
and Wilson fermions \cite{wilson}. All of them made the chiral
limit extrapolations, their results are in agreement between themselves
(see comparison Table in first reference in \cite{CDGM03})
and their average gives
\be
\im (e^2 G_E) = -(3.2\pm0.3) \, 
\left( \frac{87 {\rm MeV}}{F_0}\right)^6\,
\im \tau \, .
\ee

The result we found in \cite{epsprime} for
$\im G_8$ was
\be
\label{Q6}
\im G_8 = (4.4 \pm 2.2) \, 
\left( \frac{87 {\rm MeV}}{F_0}\right)^4\,
\im \tau 
\ee
at NLO in $1/N_c$. The uncertainty is dominated by
the quark condensate error, we used  \cite{BPR95}
\be
\label{condensate}
\langle 0 | \overline q q | 0 \rangle_{\overline{\rm MS}}
(2 {\rm GeV})
= -(0.018\pm0.004) \, {\rm GeV}^3
\ee
which agrees with the most recent sum rule determinations
of this condensate and of light quark masses --see \cite{SR}
for instance-- and the lattice light quark masses world average
\cite{WIT02}.

 We also  made a calculation for $\re G_8$ and $G_{27}$
using the technique explained  in Section \ref{motiv}
which reproduced the $\Delta I=1/2$ rule for Kaons 
within 40\% through
a very large $Q_2$ penguin-like contribution --see
\cite{DeltaI=1/2} for details. The results obtained there are
\ba
\re G_8= \left(6.0 \pm 1.7\right)\, 
\left( \frac{87 {\rm MeV}}{F_0}\right)^4\,,
\nonumber\\
{\rm and} \hskip0.5cm
G_{27}= \left(0.35 \pm 0.15 \right)\, 
\left( \frac{87 {\rm MeV}}{F_0}\right)^4\,,   
\ea
in good agreement with the results obtained
in \cite{BDP03} from a fit of $K \to \pi \pi$ and
$ K\to 3\pi$ amplitudes at NLO to data
\ba
\label{G8exp}
\re G_8 = \left(7.0 \pm 0.6\right) \, 
\left(\frac{87 {\rm MeV}}{F_0}\right)^4 \, 
\nonumber \\
{\rm and} \hskip0.5cm G_{27}= \left(0.50\pm0.06 \right) \, 
\left(\frac{87 {\rm MeV}}{F_0}\right)^4  \, .
\ea

Very recently, using a MHA to large $N_c$ QCD, the authors
of \cite{HPR03} found qualitatively similar results to those
in \cite{DeltaI=1/2,epsprime}. I.e., enhancement toward
the explanation of the $\Delta I=1/2$ rule trough large
$Q_2$ penguin-like diagrams and a matrix element of the
gluonic penguin $Q_6$ around three times the factorisable
contribution. Indications of 
large values of $\im G_8$  were  also found
in \cite{HKPS00}.

 The chiral corrections to (\ref{LOima0}), (\ref{LOrea0}),
  and (\ref{LOima2}) can be introduced   through
the $C_{\Delta I=1/2}$, $C_0$  and $C_2$  factors as follows 
\ba
\frac{\re a_0 }{\re a_2} &=& \left( \frac{\re a_0}{\re a_2}\right)^{(2)}
 \, C_{\Delta I=1/2} \, ; \nonumber \\
\frac{\im a_0}{\re a_0}
&=&\left(\frac{\im a_0}{\re a_0}\right)^{(2)} \, C_0 \, ;
\\
\frac{\im a_2}{\re a_2}
&=&\left(\frac{\im a_2}{\re a_2}\right)^{(2)} \, C_2 
+ \Omega_{\rm eff} \, \frac{\im a_0}{\re a_0} \, . \nonumber
\ea
The full isospin breaking corrections are included here
through the effective parameter $\Omega_{\rm eff}=
(0.060 \pm 0.077)$ recently calculated in \cite{CPEN03}.

We get from the fit to experimental $K \to \pi\pi$ amplitudes
in \cite{BDP03} 
\be 
C_{\Delta I=1/2} = \frac{{\cal S}_0}
{{\cal S}_2} = \frac{1.90 \pm 0.16}{1.56 \pm 0.19} = 1.22 \pm 0.15\, 
\, \, 
\ee
Where ${\cal S}_{\rm I}$ are the chiral corrections to $\re a_I$
to all orders
while we call $\cal{T}_{\rm I}$ to the chiral corrections to $\im a_I$
to all orders.   Therefore, they contain the FSI corrections
which were exhaustively studied in \cite{PPS01} plus the non-FSI
corrections which are a sizeable effect and of opposite direction.
All these chiral corrections  contain
the large overall known factor $f_K f_\pi^2 /F_0^3 \simeq 1.47$ from
wave function renormalization. 

The imaginary parts $\im a_I$ get FSI corrections identical to $\re a_I$
owing to Watson's theorem. In addition, 
both due to octet dominance in  $\re a_0$ and $\im a_0$
and to the numerical dominance of the non-analytic terms at NLO 
in $K\to \pi\pi$ amplitudes \cite{BPP98,PPS01,BDP03}
\be
C_0= \,{{\cal T}_0}/{{\cal S}_0}\,\simeq 1.0 \pm 0.2 \, ,
\ee
to a good approximation.

The situation is quite different for $\im a_2$ which is proportional to
$\im (e^2 G_E)$ at lowest order since $\im G_{27}=0$ in the
Standard Model.
{}From the works \cite{PPS01,CG02,BDP03} we also know that
\ba
\label{C2}
C_2&=&\frac{{\cal T}_2}{{\cal S}_2} 
\simeq \frac{0.70 \pm 0.21 - 0.73 L_4/10^{-3}}{1.56 \pm 0.18}
\nonumber \\
&=& 0.45 \pm0.15 - 0.47 \frac{L_4}{10^{-3}}\, . 
\ea

\begin{figure}[thb]
\begin{center}
\includegraphics[height=0.7\textwidth,angle=270]{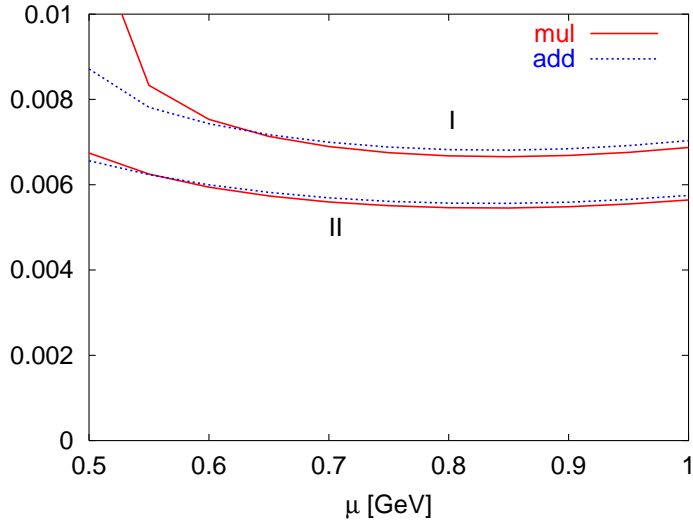}
\caption{\label{eps}  Matching of the short-distance scale
dependence of our LO in CHPT $\varepsilon'_K/\varepsilon_K$
prediction. Labels I and II are for two
different values of $\alpha_S$. The two curves
for two choices of perturbative  matching,
see \cite{epsprime}. Notice the quality of the matching.}
\end{center}
\end{figure}

 Putting all together, we get at LO in CHPT
(using (\ref{Q8}) and (\ref{Q6}))
\ba
-\frac{1}{|\varepsilon_K| \sqrt 2} \left(\frac{\re a_2}{\re a_0}
\frac{\im a_0}{\re a_0}\right)^{(2)} &\hskip-3ex=\hskip-1ex&
\hskip-1ex -(10.8 \pm 5.4) \, \im \tau \,
\,  
\nonumber\\
\frac{1}{|\varepsilon_K| \sqrt 2} \left(\frac{\re a_2}{\re a_0}
\frac{\im a_2}{\re a_2}\right)^{(2)} &\hskip-3ex=\hskip-1ex& 
(2.7 \pm 0.8 ) \, \im \tau \, .
\nonumber\\&&
\ea
And therefore, 
\ba
\label{LOeps}
&&\left(\frac{\varepsilon'_K}{\varepsilon_K}\right)^{(2)} =
((-10.8 \pm 5.4) + (2.7 \pm 0.8)) \, \im \tau \, 
\nonumber \\ 
&&=-(8.1 \pm 5.5) \,  \im \tau 
= (4.9 \pm 3.3) \times 10^{-3} \, . 
\ea
 This result is scheme independent and very stable against the
short-distance scale as can be seen in Figure \ref{eps}.
The difference with the result in Figure \ref{eps} \cite{epsprime}
is due to the new values of $\im \tau$ in (\ref{tau})
and $\re G_8$ and $G_{27}$ in (\ref{G8exp}).

Including the known and estimated higher order CHPT corrections, we get
\be
-\frac{1}{|\varepsilon_K| \sqrt 2} \, \frac{\re a_2}{\re a_0}
\, \frac{\im a_0}{\re a_0} = -(8.9 \pm 4.8) \, \im \tau \, 
\ee
and
\ba
\lefteqn{-\frac{1}{|\varepsilon_K| \sqrt 2} \, \frac{\re a_2}{\re a_0}
\, \frac{\im a_2}{\re a_2} =}&& \\ 
&&  \hskip-5ex ((1.0 \pm 0.3) + (0.5 \pm 0.7) ) 
\, \im \tau \,= (1.5 \pm 0.8) \, \im \tau \, . \nonumber
\ea
where the second part comes from isospin breaking contribution
with $\Omega_{eff} = 0.06\pm0.08$ \cite{CPEN03},
and we used $L_4=0$ in (\ref{C2}).
And therefore, 
\ba
\label{finaleps}
&&\frac{\varepsilon'_K}{\varepsilon_K} =
((-8.9 \pm 4.8) + (1.5 \pm 0.8) ) \, \im \tau \, 
\nonumber \\
&&=-(7.4 \pm 4.9) \,  \im \tau 
=(4.5 \pm 3.0) \times 10^{-3} 
\ea
to be compared to the world average \cite{NA4802,KTEV03}
\be
\label{expeps}
\re \left( \frac{\varepsilon'_K}{\varepsilon_K}\right)
{\Bigg|}_{\rm exp} = (1.66 \pm 0.16) \times 10^{-3}  \,.
\ee

Though the central value of our Standard Model
prediction in (\ref{finaleps}) is a factor around 3 
too large, within the big uncertainties it is still
compatible with the experimental result.
Two immediate consequences of the analysis above,
namely,  the LO CHPT prediction (\ref{LOeps})
 is actually very close of the
the final result (\ref{finaleps}) and second, the part with $\Delta I=1/2$
dominates  when all higher order CHPT corrections
are included.

\section{Conclusions and Prospects}

The large final uncertainty
we quote in (\ref{finaleps}) is mainly due to the uncertainties of
(1)  the chiral limit
quark condensate, which is not smaller than 20\%,
(2)  $L_5$, which is around 30\%,  and (3)
the NLO in $1/N_c$ corrections to  the matrix element
of $Q_6$, which is around 20\%.
All of them together make the present prediction for
the $\Delta I=1/2$  contribution to  $\varepsilon'_K$
to have  an error around  55\%. 
Reduction in the uncertainty of all these inputs, especially of
the quark condensate and $L_5$ is needed to obtain
a reasonable final uncertainty.

We substituted the value  used in \cite{epsprime}
 for $\im (e^2 G_E)$ 
by the one in (\ref{Q8}), notice however that numerically 
they coincide within errors.
There are also large uncertainties in the $\Delta I=3/2$
component coming from  isospin breaking \cite{CPEN03},
 and more moderate in
 the non-FSI corrections to $\im a_2$,
fortunately the impact of them in the final result is not as large as the
ones associated to the $\Delta I=1/2$ component.

 Assuming the $\Delta I=3/2$ component of $\varepsilon'_K$
  is  fixed to be the one in 
(\ref{Q8}) as indicated by  the analytic methods 
\cite{Q7Q8,CDGM03,NAR01,KPR01} and
lattice  \cite{domain,wilson}, 
one can try to extract the value of $\im G_8$ from the experimental
result in (\ref{expeps}). We get
\be
\im G_8 = (2.0 ^{+0.7}_{-0.4}) \, 
\left(\frac{87 {\rm MeV}}{F_0}\right)^4 \, \im \tau 
\ee
 i.e. the central value coincides with the large $N_c$ result
for $\im G_8$  within 30\% of uncertainty. This value has 
to be compared to our result in (\ref{Q6}).

 We have presented in \cite{BGLP03} a ladder resummation
hadronic model which keeps the good features of the ENJL model
 --CHPT at NLO, for instance--
and improves adding more short-distance QCD constraints
and  the analytic structure of large $N_c$.
 We intend to use the Green's functions calculated
within this hadronic model analytically, and  
study the relevant internal cancellations and dominant hadronic
parameters in the quantities we calculate with them. 
Work in this direction where we will study the
origin of the large chiral corrections to $\hat B_K$ 
we found in \cite{BK} is
in progress \cite{BGLP03b}.  This program will also be  extended
to study  the origin of the large value for $\im G_8$ in
(\ref{Q6}) which we got in \cite{epsprime}.
Large values of $\im G_8$ were previously pointed out
in \cite{HKPS00} and more recently in \cite{HPR03}.

\vspace*{-0.3cm}
\section*{Acknowledgments}
It is a pleasure to thank Stephan Narison for the invitation
to this very enjoyable conference. We also thank
Vincenzo Cirigliano for comments.
E.G. is indebted
to MECD (Spain)  for a F.P.U. fellowship. J.P. also thanks the hospitality
 of  the Department of Theoretical Physics at Lund University where
 this work was written.

\end{document}